\documentstyle[11pt,paspconf,epsf]{article}

\def\spp{\sigma_\pi/\pi}
\def\sp{\sigma_\pi}
\def\infig#1#2{\epsfxsize=#2cm \centering{\mbox{\epsfbox{#1}}}\vspace{-0.4cm}}
\def\FC{FC}
\def\S{SZ}
\def\MF{MF}
\def\OG{OGS}
\def\L{LGT}

\begin{document}

\title{The Cepheid Distance Scale after Hipparcos}
\author{Frédéric Pont}
\affil{Observatoire de Genève, Switzerland}

\begin{abstract}

More than two hundred classical cepheids were measured by the Hipparcos astrometric satellite, making possible a geometrical
calibration of the cepheid distance scale.  However, the large average
distance of even the nearest cepheids measured by Hipparcos implies trigonometric parallaxes of at most a few {\it mas}. Determining unbiased distances and absolute magnitudes from such high relative error parallax data is not a trivial problem.

In 1997, Feast \& Catchpole announced that Hipparcos cepheid parallaxes indicated
a Period-Luminosity scale 0.2 mag brighter than previous calibrations, with important
consequences on the whole cosmic distance scale. In the wake of this initial study, several authors have reconsidered the question, and favour fainter calibrations of
cepheid luminositites, compatible with pre-Hipparcos values.

All authors used equivalent data sets, and the bulk of the
difference in the results arises from the statistical treatment of the
parallax data. We have attempted to repeat the analyses of all these studies and test them with
Monte Carlo simulations and synthetic samples. We conclude that the initial Feast \&
Catchpole study is sound, and that the subsequent studies are subjected
in several different ways to biases involved in the treatment of high relative
error parallax data. We consider the source of these biases in some detail. We also propose a reappraisal of the error budget in the final Hipparcos
cepheid result, leading to a PL relation -- adapted from Feast \& Catchpole -- of
\[
M_V=-2.81 ({\rm assumed}) \log P -1.43\  \pm 0.16 ({\rm stat}) \ \,^{+0}_{-0.03} ({\rm syst})
\]
We compare this calibration to recent values from cluster cepheids or the surface brightness method, and find that the overall agreement is good within the uncertainties.

We conclude by commenting on the mismatch between the cepheid parallax distance scale and kinematical determinations, for cepheids as well as RR Lyrae.

\end{abstract}


\section{Introduction}

The cepheid distance scale is the central link of the cosmic distance
ladder. Because even the nearest cepheids are more than 100 pc away
from the Sun (Polaris at $\sim 130$ pc, $\delta$ Cep at $\sim 300$ pc),
 no reliable parallax measurements were available for
cepheids before the Hipparcos mission. The zero-point of the
Period-Luminosity (PL) relation was calibrated by secondary
methods, using cepheids in open clusters and associations or Baade-Wesselink techniques.
The Hipparcos data for cepheids opened the possibility of a geometric
determination of the zero-point of the PL relation.

Soon after the Hipparcos data release, Feast \& Catchpole (1997, \FC )
announced that Hipparcos cepheid data indicated a zero-point about 0.2
mag brighter than previously thought (implying $\mu$=18.70 for the LMC). Adopting the FC notation for
the PL relation :
\[
M_V=\delta \log P + \rho
\]
they find $\rho=-1.43\pm 0.13$ (for $\delta=-$2.81 assumed).
This result attracted considerable attention, since it implied a $\sim$10\%
downward revision of the value of $H_0$ obtained from galaxy recession
velocities. In turn, the higher expansion ages derived could become
compatible with  the new, lower ages  for globular clusters obtained
from Hipparcos subdwarfs, resolving the ``cosmic age problem'' (the fact
that globular cluster ages were found to be much older than the
expansion age of the Universe).

\begin{figure}[thb]
\infig{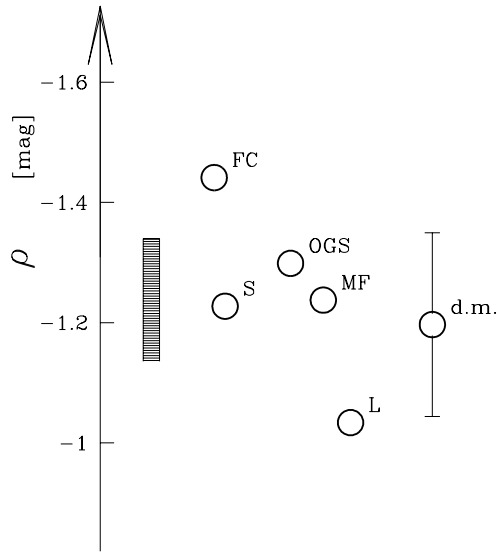}{8.8}
\caption{Calibrations of the cepheid Period-Luminosity relation zero-point ("$\rho$") from Hipparcos parallaxes : Feast \& Catchpole (1997), Szabados (1997), Oudmaijer et al. (1998), Madore \& Freedman (1998), "direct method" (see Par.~\ref{dm}), and Luri et al. (1998). Typical error bars indicated for one calibration. The dashed area indicates a typical range of pre-Hipparcos values. The PL relation slope ("$\delta$") is fixed at $-$2.81. See also Fig.~\ref{concl}. }
\label{all}
\end{figure}

In the past year however, several authors reconsidered the Hipparcos
cepheid data, and voiced criticism against the \FC\ result. All
subsequent studies obtain calibrations for the zero-point of the PL
relation similar to pre-Hipparcos values, or fainter (see sketch Fig.~\ref{all}).
Szabados (1997) pointed out that known or suspected binaries were
abundant in the cepheid sample, and that the PL relation found after
removing them was compatible with pre-Hipparcos values. Madore \&
Freedman (1998) repeated the whole analysis with multi-wavelength data
(BVIJHK). They also concluded on a fainter zero-point. Oudmaijer et al.
(1998), in a study devoted to the effect of the Lutz-Kelker bias on
Hipparcos luminosity calibrations, analyse the cepheid data as
an illustration and derive $\rho=-1.29 \pm 0.02$. Finally, Luri et al.
(1998) apply the ``LM method'' (Luri et al. 1996) to the cepheid sample and
find a very faint zero-point of $\rho=-1.05 \pm 0.17$.

The situation,  one year after the Hipparcos data
release, is therefore very perplexing : while before Hipparcos the zero-point of the cepheid PL relation seemed reasonably well determined within 0.1 mag (e.g. Gieren et al. 1998), values derived from the same Hipparcos parallax data cover a range of 0.4 mag!

Fortunately, this distressing situation may be only temporary. As all the studies considered
use the same parallax data and similar photometric and
reddening values for Hipparcos cepheids, the differences are
almost entirely due to the statistical procedures used in the analyses.
Contrarily, for instance, to the case of the very metal-poor globular
cluster distance scale where observational uncertainties may remain the
limiting factor (see review by Cacciari in this volume), it may be
hoped that the disagreements about the PL zero-point can be resolved by
evaluating the different approaches.

This is what we are attempting here. From statistical
considerations on one hand, Monte Carlo simulations of the different
procedures on the other hand, we have tested the robustness and possible
biases of the different procedures. The results are presented below in Par. 3 to 8, study by study, beginning with an imaginary author using the unsophisticated "direct method". It appears that the difficulties involved in the treatment of high relative error parallax data are at the core of the question. On this subject, the review by Arenou in this volume constitute an excellent complement to the present contribution -- showing the pitfalls easiest to overlook. 
As far as cepheids are concerned, we shall contend here that, in fact, the situation is rather clear, and
that much of the apparent disagreement is caused by 
statistical biases that have already been reported in  dealing with parallax data.

\section{Hipparcos cepheids}

Around 200 cepheids were measured with Hipparcos. All but the nearest
ones are so remote that $\pi_{real}<\sigma_\pi$\footnote{Parallaxes are noted $\pi$ and their uncertainty $\sigma_\pi$}. There are 19 of them with $\spp<50$\%. The closest
cepheid, Polaris, was measured at $\pi=7.56\pm 0.48$ {\it mas}. $\delta$ Cep itself
has $\pi=3.32\pm 0.58$ {\it mas}. Despite the low accuracy of individual parallax
determinations, it is not unreasonable to attempt an accurate
determination of the PL relation zero-point, because {\it relative}
distances are precisely known through the PL and PLC relations. Each individual parallax can therefore be seen as a measurement of the PL zero-point. A large number of low-accuracy parallaxes can yield a reliable combined value of the zero-point.

\section{Dangers of the ``direct method''}

\label{dm}
The most natural way to infer absolute magnitude from parallax would be to
calculate distances from the inverse of the parallax and use Pogson's law :
\begin{equation}
m_V-M_V= 5 \log (1/\pi)-5+a_V
\label{pogson}
\end{equation}
Then, the zero-point of the PL relation could be fit in the Period-$M_V$
plane, by some kind of weighted least-square.

\begin{figure}[htb]
\vspace{-5cm}
\infig{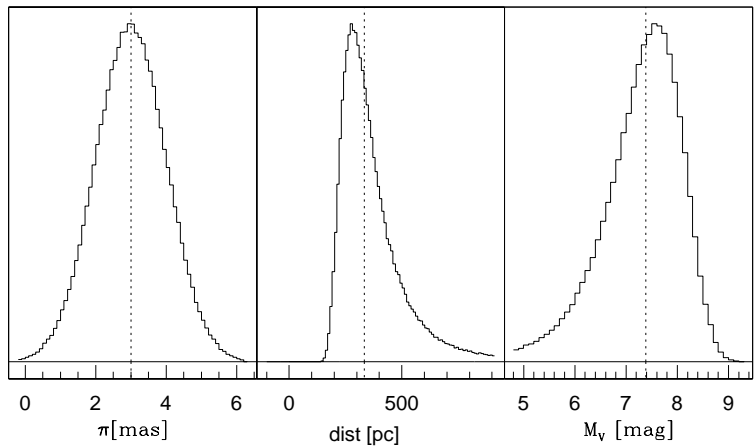}{12}
\caption{Statistical distributions of the parallax, distance computed from $d=1/\pi$ and absolute magnitude from Equ.~\ref{pogson} with $\pi_{real}$=3 mag and $\sp$=1 mas. The distance and absolute magnitude distributions are biased and asymmetrical. }
\label{trois}
\end{figure}

In fact, this is an {\it extremely} biased way to proceed when the
relative errors on parallax are high ($\spp >\sim $20\%). Fig.~\ref{trois}
shows graphically why the inverse of the parallax is a biased estimator
of the distance, and how the magnitude derived from Equ.~\ref{pogson} has a very
asymmetrical and skewed distribution if the statistical distribution of
the parallax is gaussian. This effect was pointed out by several
authors in the wake of the publication of the Hipparcos catalogue (e.g. Luri \& Arenou 1997). An essential condition to the
use of least-square fit is the symmetry of the error distribution, and
if it is not satisfied, first-order biases are to be expected.
Moreover, in order to use Equ.~\ref{pogson}, negative parallaxes must be ignored and the data selected by some cut in $\spp$. Both selection criteria introduce further biases. By discriminating against low $\pi_{meas}$ at a
given $\sp$ and $\pi_{real}$, they bias the result towards lower
distances and fainter magnitudes.

Monte Carlo simulations with samples similar to the actual Hipparcos
cepheid sample show that this ``direct method'' leads to a bias of $\sim$0.2
mag towards a fainter PL relation zero-point. An important bias indeed, due
exclusively to the incorrect statistical treatment of parallax data.

\section{Shifting to parallax space : Feast \& Catchpole 1997}

\label{pfc}

 FC avoid the difficulties of the $\pi \!\rightarrow\! M_V$
transformation with a change of variable that allows the parallaxes to
be combined linearly. Instead of deriving the PL relation zero-point
from $M_V=\delta \log P + \rho$ and Equ.~\ref{pogson}, FC compute $10^{0.2\rho}$ from the mathematically equivalent relation
\begin{equation}
10^{0.2\rho}=0.01 \ \pi\ 10^{0.2(m_V-a_V-\delta \log P)}
\label{fc}
\end{equation}
The final value of $\rho$ is recovered from the average of the values of
$10^{0.2\rho}$ weighted by the uncertainty on the right term of~(\ref{fc}).

\begin{table}[b]

\begin{tabular}{l l l l}
\centering
Sample & $\rho$ & $\sigma_\rho$ & N \\ \hline
all (27 stars) & $-$1.426 & 0.128 & 1000\\
without $\alpha$ Umi (26 stars)& $-$1.433 & 0.197 & 1000\\
without overtones (20 stars) & $-$1.436 & 0.232 & 1000 \\
weights $<$ 50 (5 stars) & $-$1.448 & 0.267 & 1000 \\ \hline
\end{tabular}
\caption{First test of the FC method on synthetic samples. 
Starting from samples identical to the real one, each Monte Carlo 
realization varies the errors on the parallaxes according to the stated $\sigma_\pi$.
The input value of $\rho$ is $-$1.43 in all cases. }
\label{fc1}
\end{table}

\begin{table}[t]
\begin{tabular}{l l l l}
Number of stars & $<\!\rho\!>$ & $\Delta M_V$ & limit weight \\ \hline
50'000& $-$1.431 & 0.2 & 10 \\
50'000 & $-$1.438 & 0.5 & 10 \\
50'000 & $-$1.461 & 1.0 & 10 \\
20'000 &  $-$1.430 & 0.2 & 50 \\ \hline
\end{tabular}
\caption{Second test of the FC method on synthetic samples. Samples of cepheids filling a volume of space larger than the selection limits are built. The FC procedure is applied to the subsample fulfilling the FC selection criteria. This simulation takes the classic Lutz-Kelker bias into account. The  $\Delta M_V$ parameter is the width of the instability strip. The input $\rho$ is $-$1.43 in all cases.}
\label{fc2}
\end{table}

At first sight the procedure adopted by FC looks unnecessarily complicated.
But we saw in the previous section why a more straightforward approach
is not preferable. The FC method removes the statistical biases affecting the direct method: since the parallax appears linearly in~(\ref{fc}), negative
parallaxes can be kept, no $\spp$ cut is needed, and the uncertainties are
symmetrical. However, the condition for the use of this method is that
the uncertainties on the exponent $0.2(m_V-a_V-\delta \log P)$ be smaller than
the uncertainties on $\pi$. For this reason, the method is only reliable for a
group of objects with errors on the relative distances much smaller than the parallax errors, such as the
Hipparcos cepheids. Any dispersion of the exponent $0.2(m_V-a_V-\delta \log P)$ will
make the distribution of errors on the right term of~(\ref{fc}) asymmetrical again, and result in a bias towards
brighter magnitudes.

We have tested the FC procedure with Monte Carlo simulations on synthetic
samples of various composition.  The effect of
modifying several assumptions was tested, such as varying the slope
of the PL relation, the width of the instability strip or the spatial
distribution of cepheids. Samples were also drawn from a larger volume, so that
classic Lutz-Kelker biases would be modeled.\pagebreak \ Representative results are shown on Tables~\ref{fc1} and~\ref{fc2}. The
conclusion is that the FC method is sound and robust, and that
systematic biases are smaller than 0.03 mag\footnote{Similar results were obtained by X. Luri (priv.
comm.) who has also extensively tested the FC procedure. These small residual biases are caused by the asymmetrical effect of the dispersion in the exponent $0.2(V_0-\delta \log P)$ of Equ.~\ref{fc}.}.

However, the {\it dispersion} of the results recovered in the
simulations is substantially higher than that stated in FC, indicating
that the final uncertainty may have been underestimated. We return on this
point in Par.~\ref{error}

\section{Possible effect of binaries : Szabados 1997}

\begin{figure}
\infig{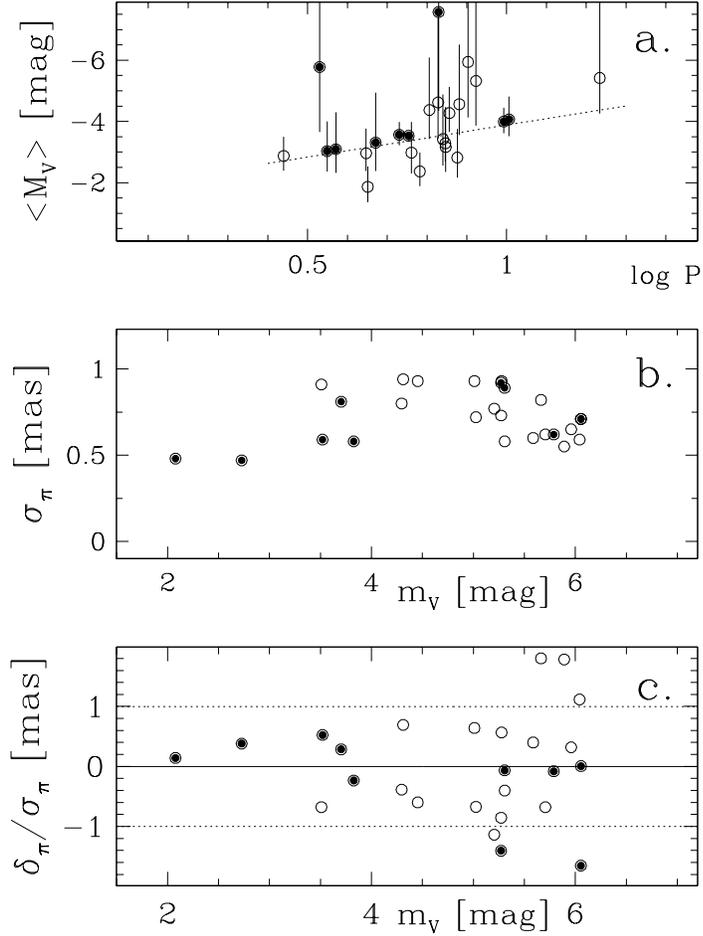}{9.8}
\caption{{\bf a.} Period-luminosity diagram for the nearest Hipparcos cepheids as in Szabados (1997). Known or suspected binaries in white. Error bars from the parallax uncertainties. {\bf b.} Parallax uncertainties $\sp$ as a function of apparent magnitude. {\bf c} Normalised parallax residuals $(\pi_{observed}-\pi_{PL})/\sp$ as a function of apparent magnitude. $\pi_{PL}$ is the parallax expected with the distance computed from the PL relation. The distribution of parallax residuals, for suspected binaries as well as for single stars, is compatible with a normal distribution.}
\label{sz}
\end{figure}

A large fraction of known cepheids are confirmed or suspected binaries.
Szabados (1997, \S) has pointed out that  binary cepheids showed more scatter in the PL diagram than single cepheids, and suggested that this could be due to the noise induced by binarity on Hipparcos parallax determinations. As an orbit of $\sim$1~AU has the amplitude of the parallax at any distance, unrecognized companions could in principle interfere with the parallax measurement. By fitting a PL relation on single cepheids only (Fig.~\ref{sz}a), \S\ recovers a zero-point equivalent to the pre-Hipparcos values (about 0.2~mag fainter than FC).

As shown by Fig~\ref{sz}a, suspected binaries have larger error bars on average than single cepheids. 
But given these error bars, the scatter for binary suspects is compatible with the uncertainties (as confirmed by a Kolmogorov-Smirnov test). The question is to understand why binaries have larger error bars, and for this the $\log P-M_V$ plane is not a good representation, as explained in Par.~3, because the same {\it parallax} uncertainty can result in greatly different $M_V$ uncertainties.

When the uncertainties are considered in parallax space (Fig~\ref{sz}b./\ref{sz}c.), it appears that at a given magnitude the binary cepheids do not have significantly larger parallax uncertainties than the single cepheids. The high dispersion of the binary group on Fig.~\ref{sz}a is simply due to the fact that, on average, suspected binaries are fainter and more remote. Now, this has to be a chance effect due to low-number statistics (unless some weird mechanism weeds out binary cepheids off the solar neighbourhood). 

Another aspect of Fig~\ref{sz}a is that eight single stars look much nearer to the mean relation than their uncertainties would indicate, adding to the visual impression that binary suspects are much more scattered. These stars have $\sp$=0.5-0.7 mas, and much smaller residuals : $<\!\pi_{obs}-\pi_{PL}\!>$=0.07 mas. Again, this can only be due to chance. The alternative is to suggest that, for some reason, Hipparcos parallaxes are a factor 7-10 more precise for single cepheids than for any other star in the catalogue, a rather unreasonable hypothesis. If this sounds like a strange coincidence, one should keep in mind that the cepheid sample was split into two parts under several criteria to check for systematic effects -- single and binary, overtone and fundamental pulsator, low and high period, low and high reddening -- and that a slightly strange-looking distribution for 8 points according to one of these criteria should not be over-interpreted. A Kolmogorov-Smirnov test indicates that the normalised parallax residuals $(\pi_{obs}-\pi_{PL})/\sp$ (see Fig.~\ref{sz}c) for binaries only, for single stars only and for the combined sample, are all compatible with a normal distribution. The lowest KS coefficient, that for single stars, is 0.27.

Therefore, there is no statistically significant indication that the suspected binary cepheids suffer some additional noise on the Hipparcos parallax measurements.

Does the exclusion of suspected binaries change the zero-point derived from Hipparcos parallaxes? \S\ uses a straightforward fit in magnitude space to calculate the PL relation of single cepheids. We saw in Par.~3 how biased the results could become. In fact, when analysed with the procedure used by FC, the single cepheids only, the binaries only and the whole sample give  similar results ($\rho=-$1.51, $-$1.36 and $-$1.43 respectively). The zero-point $\sim$0.2 mag fainter found by \S\ is not due to the exclusion of suspected binaries, but to the biases caused by the use of magnitudes calculated from parallaxes.

The implications are that :\\
1- There is no statistically significant indication that binarity affects Hipparcos parallaxes for cepheids. \\
2- Keeping or removing the suspected binaries gives essentially the same result for the PL relation zero-point.\\

\section{Multi-wavelength magnitude analysis : Madore\&Freedman 1997}

Madore \& Freedman (1998, \MF) reconsider the calibration of
Hipparcos cepheids, using data in several visible and infrared
wavelengths (BVIJHK) rather than the traditional B and V. This reduces the
number of objects available, as only 7 Hipparcos cepheids have been
measured in all six wavelengths.
\MF\ compute magnitudes from the standard formula
\begin{equation}
m_i-M_i=5 \log (1/\pi)-5 +a_i 
\label{emf}
\end{equation}
and calculate the PL zero-point $\rho$ by averaging the magnitude residuals with
weights $\omega\equiv \pi^2/\sigma_\pi^2$.
 This procedure yields  values for  $\rho$ that depend significantly on wavelength
(see column two of Table~\ref{tmf}), a dependence that \MF\ attribute to
reddening problems
\footnote{\MF\ take individual reddenings from the Fernie et al. (1995) catalogue (and not from the reference given in the article, Fernie, Kamper \& Seager 1993, that does not contain reddenings). \MF\ state that they use the same unreddening procedure as \FC. However, Fernie et al. use mulicolour calibrations for reddenings, whereas \FC\ calculate reddenings from a mean Period-Colour (PC) relation. Thus \MF\ do not benefit from the compensating effect of combining PC reddenings with a PL relation (see
for instance Pont et al. 1997 or \FC). As a consequence, the biases are amplified, which may explain part of the dependence of their zero-point on wavelength.}. At the longer wavelengths, \MF\ recover the values
found by FC, whereas in the infrared fainter values are found, corresponding to the pre-Hipparcos calibration and $\mu_{LMC}\sim 18.5$ mag.

\begin{table}
\centering
\begin{tabular}{l | l l || l  |}
Filter & $\mu_{LMC}$ & $\Delta \rho$ & $\Delta' \rho$\\ \hline
B & 18.74$\pm$0.36 & $-$0.04 & \\
V & 18.67$\pm$0.24 & +0.03 & $-$0.22 \\
I & 18.71$\pm$0.20 & $-$0.01 &  \\
J & 18.44$\pm$0.24 & +0.26 & $-$0.26 \\
K & 18.57$\pm$0.14 & +0.13 & $-$0.15 \\\hline
\end{tabular} 
\caption{Column 2 and 3 : values of $\mu_{LMC}$ recovered by MF in the multi-wavelength analysis, with the offset compared to the value of FC. Column 4 : offset between the V, J and K calibrations of Laney \& Stobie (1994) and the \FC\ method applied to the \MF\ data. In this case, the brighter Hipparcos cepheid scale is recovered and the strong wavelength dependence vanishes.}
\label{tmf}
\end{table}

It should now be clear that the \MF\ procedure is subject to very large
biases, as it corresponds to the type of approach outlined in Par.~3 above. 
The $M_i$ calculated from Equ.~\ref{emf} are very biased estimators of the
absolute magnitude for high values of $\spp$. The introduction of weights depending
on the observed $\spp$ ratio only makes matters worse. The effect may be illustrated
 with a simple example : consider 3 cepheids at
the same distance, say 500 pc,  measured by Hipparcos with
$\sp$ = 1 mas. The real $\pi$ is 2 mas (1/500 pc), and let the measured
$\pi$ be 1, 2 and 3 mas respectively. Table~\ref{3cep} shows the magnitudes
derived for these 3 objects from the measured parallaxes using Equ.~\ref{emf} and the weights from \MF. A weighted average gives a calibration that is 0.46 mag
too faint! A better procedure would be to work in parallax space, in this
case averaging the parallaxes weighted by $\sigma_\pi$, to obtain the
{\it parallax} of the zero-point, a procedure similar in essence to \FC.

\begin{table}[h]
\centering
\begin{tabular}{ | l l l l| }
dist &$\pi_{meas}$ & $M_V residual$ & weight \\ \hline
500 pc & 1 mas & $-$1.5 mag & 1 \\
500 pc & 2 mas & 0 mag & 4 \\
500 pc & 3 mas & +0.88 mag & 9\\ \hline
\multicolumn{2}{| l}{Weighted mean :} & +0.46 mag & \\ \hline
\end{tabular}
\caption{"3 objects" illustration of the bias affecting the \MF\ procedure. Three cepheids at 500 pc are measured in parallax with a typical error of 1 mas. Magnitudes are computed from the parallax with Equ.~\ref{emf} and weights from $(\pi/\sigma_\pi)^2$. In this case the \MF\ procedure results in a 0.46 mag bias towards fainter magnitudes.}
\label{3cep}
\end{table}

The biases affecting the \MF\ results were also checked by applying the \FC\ method to the \MF\ multi-wavelength data. The resulting zero-point was compared in V, J and K to the Laney \& Stobie (1994) calibrations (Column~4 of Table~3). 
The resulting zero-point is coherent at all three wavelengths, $\sim$ 0.2 mag brighter than Laney \& Stobie, and
corresponds to the ``bright'' \FC\ calibration. 

Biases are also apparent when the \MF\ procedure is applied to synthetic samples.  As expected, the recovered luminosity zero-point is systematically too faint, on average 0.25 mag too faint with synthetic samples similar in distance distribution to the actual sample.

Thus the disagreement of \MF\ with \FC\ is
not due to the use of multi-wavelength data, but can be entirely attributed to the treatment of parallax data. This confirms the necessity of carefully considering the subtleties involved
in deriving magnitude calibrations from high $\spp$ parallax (see Brown et al. 1997, Luri \& Arenou 1997).

\section{Presence of the Lutz-Kelker bias : Oudmaijer et al. 1998}

\begin{figure}
\infig{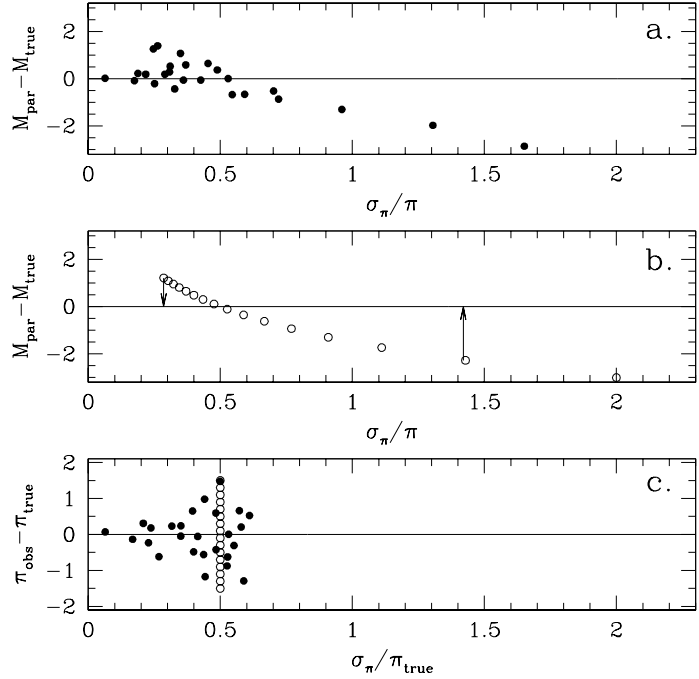}{9.8}
\caption{{\bf a.} $M_{\rm par}-M_{\rm true}$ as a function of $\spp$ for the 26 cepheids used by \OG. 
$M_{\rm par}$ is the $V$ magnitude computed from the parallax with Equ.~\ref{pogson}, $M_{\rm true}$ 
the magnitude from the PL relation. {\bf b.} Position of an object of fixed true parallax and magnitude 
when the observed parallax varies of $\pm 1.5\sp $, showing the correlation of the abscissa and the ordinate. The arrows illustrate the correction of Equ.~\ref{corog}.
{\bf c.} {\it Parallax} residuals ($\pi_{\rm obs}-\pi_{\rm true}$) as a function of $\sigma_\pi/\pi_{true}$ 
for the 26 cepheids and the imaginary object as in (b.). $\pi_{true}$ is the parallax expected from the PL 
relation distance.\hspace{4.5in}
The feature observed in the upper and middle panels is only due to the correlated nature of the abscissa and the ordinate. On the uncorrelated last panel, a correction such as Equ.~\ref{corog} becomes unnecessary.}
\label{fog}
\end{figure}

The Oudmaijer et al. (1998, \OG) study is devoted to showing the
presence of the Lutz-Kelker and Malmquist biases in Hipparcos data, and calculating statistical corrections on individual measurements to
compensate for these biases.
\OG\ consider the plot of the magnitude residual $\Delta M_V$ versus $\spp$
(see Figure~\ref{fog}a for the cepheids) as evidence for the presence of such biases. 
The dependence of  $\Delta M_V$ on $\spp$ is indeed impressive.

But this interpretation is incorrect -- as detailed by Arenou at this
meeting: the features in Fig~\ref{fog}a are primarily due to the fact that the abscissa and ordinate are heavily
correlated. Let us suppose that $\sigma_\pi$ is a constant, as is nearly
the case for Hipparcos parallaxes, then the abscissa is $\spp \sim 1/\pi$, while the
ordinate is
\[
\Delta M_V \equiv M_V^{par}-M_V^{true} = 5 \log(1/\pi)-5-M_V(true) \sim -log(\pi)
\]
Both axes strongly depend on the same measured parallax $\pi$, and the relation
observed in Fig.~\ref{fog}a (and Fig. 2, 3 and 4 of \OG) only
reflects this direct correlation, and does not as such reveal any bias.

Fig.~\ref{fog}a would be more useful if the abscissa contained the {\it real} $\spp$, and
not the {\it observed} $\spp$. Unfortunately the real $\pi$ is unknown. For
a given real $\pi$, the variations of the observed $\pi_{obs}$ due to parallax
uncertainties affects $\sigma_\pi/\pi_{obs}$ and $\Delta M_V$ in a correlated way and move
data points along diagonal lines in the diagram, as illustrated in Fig.~\ref{fog}b. 
$\spp_{obs}$ is a reliable indication of $\spp_{real}$ only if $\sp<\!<\pi$, which is not the case for cepheids.

\OG\ propose the following correction on individual data
\begin{equation}
\delta M = \left\{ 1- \left[ \frac{\sigma^2_{M_0}}{\sigma^2_{M_0}+4.715 (\sigma_\pi/\pi)^2}\right] \right\} (M_0-M_{obs})
\label{corog}
\end{equation}
where $M_0$ is the expected magnitude (the ``true'' magnitude).
As this correction depends on the {\it unknown true magnitude $M_0$}, it
is not determined unless a true magnitude $M_0$ is assumed.  In the case
of cepheids, \OG\ chose the procedure of trying different values of
$\rho$ until the residuals of the magnitudes corrected by Equ.~\ref{corog} reach a
minimum, and recover a value of $\rho$ similar again to the pre-Hipparcos
calibrations. Their procedure is illustrated by the arrows in Fig.~\ref{fog}b: a
large correction depending on $M_0$ and $\spp$ is applied (Equ.~\ref{corog}), and
different $\rho$ are tried until the residuals are minimum. \OG\ take as a
proof that their procedure has corrected for the biases the fact that,
after correction, the $\spp$ vs. $\Delta M_V$ plot looks very tidy. A closer
look at Equ.~\ref{corog} shows that this has nothing to do with bias correction :
because the correction itself tends to ($M_0-M_{obs} $) as $\spp$ becomes
high, the corrected $M_{obs}$ is forced to $M_0$ ($M_{obs}+\delta M\simeq M_{obs}+(M_0-M_{obs}) = M_0$), so that the residuals
artificially tend to zero! If an abscissa independent of the ordinate is chosen, e.g. $\sigma_\pi/\pi_{true}$, where $\pi_{true}$ is the parallax expected from the PL relation, the largest part of the apparent bias vanishes (Fig.~\ref{fog}c), and correction Equ.~\ref{corog} becomes unnecessary\footnote{The remaining part of the bias is the much smaller classical Lutz-Kelker bias, due to the fact that different parallax intervals cover very different space volumes. In the case of Hipparcos cepheids selected as in \FC, it amounts to $\sim$0.02 mag}.

We have tested the \OG\ procedure for cepheids on synthetic samples and
find that it gives systematically too faint results by $\sim$0.17 mag. It also
fails the ``3 cepheid'' test of the previous section (giving a bias of
0.36 mag). The crux of the matter here is that, as the relative error on
parallax increases, the {\it measured} $\spp$ becomes an increasingly bad
estimator of the {\it real} $\spp$. In our small ``3 cepheid'' example,
the measured $\spp$ is 33\%, 50\% and 100\%, while the true $\spp$ is 50\%. As a rule of thumb,
$\spp$ should no longer be used, even in statistical corrections, for
values higher than about 20\%.

Due to its indirect nature (the true magnitude must first be assumed and then corrected residuals are minimised), the \OG\ method is also quite unstable. One can get a feeling of this instability by
comparing the results of the \OG\ method with that of the \FC\ method on several cepheid subsamples (Table~\ref{tog}).

\begin{table}
\begin{tabular}{l l l}
Sample & $\rho$ with \OG\ method& $\rho$ with \FC\ method\\ \hline
26 stars & $-$1.34 & $-$1.42 \\
without $\alpha UMi$ & $-$1.16 & $-$1.44 \\
without overtones & $-$1.29 & $-$1.49 \\
without binaries & $-$1.35 & $-$1.36 \\
only binaries & $-$0.84 & $-$1.51 \\ \hline
\end{tabular}
\caption{The \OG\ and \FC\ procedure compared on several subsamples of the Hipparcos cepheid data. Note the instability of the \OG\ results.}
\label{tog}
\end{table}

We conclude that while the \OG\ method might be justified to estimate the absolute magnitude of {\it individual} objects from a high $\Delta M_V$ population with low $\spp$ and about which nothing is known, it is far from optimal in the case of the
cepheids, which have high $\spp$ and low $\Delta M_V$. In that case, $\pi_{true}$ is
better known from the PL relation itself than from $\pi_{meas}$, and the use
of $\spp$ can be avoided by working in parallax space.

\section{Maximum likelihood method : Luri et al. 1998}

\label{lu}
Luri et al. (1996) have devised a maximum-likelihood method of
determination of absolute magnitudes from Hipparcos data that takes all
available data into account, including proper motion and radial
velocity. Luri et al. (1998, \L) apply this method to the zero-point of the cepheid PL relation and derive $\rho$=1.05$\pm 0.17$ mag for a fixed slope of $\delta\equiv -2.81$, a
value 0.38 mag fainter than \FC\ using exactly the same sample.

First, it should be noted that \L\ use the global reddening model of
Arenou et al. (1992), that gives reddenings on average 0.05 mag higher than
those usually adopted for cepheids. If the reddening scale of \FC\ is
adopted instead, the result of the LM method becomes $\rho=-$0.89 mag, an even fainter value.

In collaboration with X. Luri, we tested the method with synthetic
samples, and no significant bias was found. The key to this puzzling
problem was pointed out by F. van Leeuwen at this meeting : further
tests showed that the \L\ solution is much more sensitive to kinematical
data (proper motions and radial velocities) that to the parallaxes. In fact, to the first order,
the parallaxes have no influence on the solution, so that the LM method
becomes similar in principle to a statistical parallax analysis. It
cannot be directly compared to the geometrical distance determinations considered
above.

The large disagreement of the LM method with \FC\ remains a question
that deserves detailed study, as it may contain precious hints on
cepheid distances or kinematics (see Par~12 below).

\section{A note on overtone pulsators}

Overtone pulsators, cepheids that pulsate in the first harmonic rather
than the fundamental mode, are usually identified by their low amplitude
and sinusoidal light curve. Their period is about 30\% shorter than a
fundamental pulsator of the same luminosity. In a luminosity calibration, detected overtones can
either be removed or their period adjusted by 30\%, but the presence of
undetected overtones may bias the result.

The possible presence of undetected overtones was tested by repeating the zero-point determination on subsamples selected by period intervals. Because overtones usually have small periods, their undetected presence should appear as a period dependence of the solution. No significant trend was found, but the sample is not large enough to exclude such a trend below the $\sim$0.1 mag level.

\pagebreak

\section{Value and uncertainty of the Hipparcos cepheid PL zero-point}

\begin{figure}[hbt]
\infig{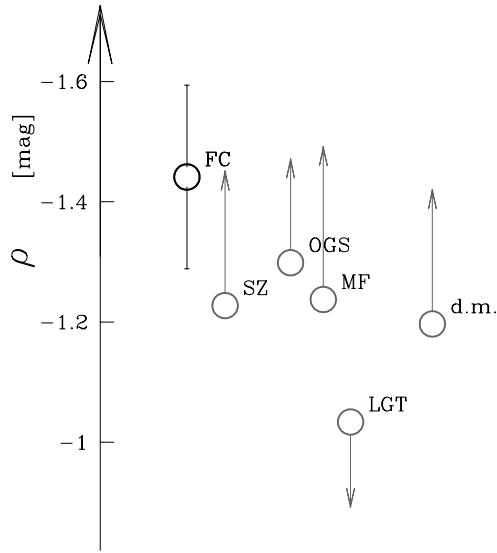}{8.8}
\caption{Calibrations of the cepheid PL relation zero-point as in Fig~\ref{all}, with the bias corrections discussed in this paper. The error bars of \FC\ are increased. The other results are modified to account for the biases in the treatment of parallaxes, except \L, for which the reddening scale has been adjusted.}
\label{concl}
\end{figure}

Fig.~\ref{concl} graphically summarizes our conclusions : the Hipparcos
parallaxes for cepheids do indeed indicate a magnitude calibration
brighter than previously accepted ($\rho=-1.43 \pm 0.16$ for
a fixed PL slope  $\delta\equiv -$2.81), as found by \FC. 

All other subsequent analyses that we considered suffer from strong systematic biases due
to the procedures used to infer magnitude calibrations from parallax
data
\footnote{With the interesting exception of \L, that, as explained in Par.~\ref{lu}, cannot be strictly considered as a parallax calibration, but should rather be seen as a kinematical calibration.}.
 The impression one could infer by ``weight of numbers'',  from an
uncritical list of all calibrations, namely that after all Hipparcos cepheid
data confirm previous  magnitude calibrations, is therefore
misleading. On the contrary, we confirm the conclusion by FC that a brighter cepheid PL calibration is implied by
the Hipparcos parallaxes.

Let us now consider the {\it uncertainty} on this value. In our simulations,
the dispersion in the recovered $\rho$ came out 15\% to 50\% higher than
the uncertainties derived in FC. In fact, the error in FC is
lower that the uncertainty caused by the propagation of Hipparcos parallax
uncertainties alone. This is due to the fact that FC calculate the
uncertainty from the residuals, not from the Hipparcos $\sp$, and that the residuals are on average
lower than the uncertainties:
\[
<\!\frac{\pi_{observed}-\pi_{PL}}{\sigma_\pi}\!> =0.87
\]
Obtaining normalised residuals lower than unity can have two causes:
either the uncertainties were overestimated, 
or the effect is due to chance and low-number statistics.

It is unlikely that Hipparcos parallax errors were overestimated for
cepheids, and we shall rather assume that a statistical fluctuation causes the residuals to be smaller than the uncertainties. In that case, the final error
to be used is not the one from the residuals, but the one propagated from
the uncertainties on the parallaxes. With this consideration, we modify
the error in FC upwards to 0.16 mag, noting that this uncertainty is due only to the $\sp$ and than any other source of uncertainty (e.g. reddening scale, PL slope) would be additional. Our Monte Carlo simulations (Table~\ref{fc1}) show typical scatters as high as 0.20 mag in the recovered $\rho$ for the sample without $\alpha$ UMi.

Recalling the discussion of section~4, we add a possible systematic
bias of $\ ^{+0.03}_{-0}$ mag, for a final Hipparcos cepheid PL relation zero-point modified from FC of
\begin{equation}
M_V=-2.81 \log P \underline{-1.43 \pm 0.16 \,[{\rm stat}]\, \, ^{+0.03}_{-0} \,[{\rm syst}]\,} 
\end{equation}

If our rediscussion of the uncertainties of the FC result is
correct, the Hipparcos calibration is not incompatible with
previous calibrations  from cluster cepheids or from surface brightness
techniques. The uncertainty on the geometrical calibration remains high
and does not force a shift of the distance scale. The Hipparcos parallax
data do however indicate that the real PL relation is probably situated
near the bright end of previous uncertainty intervals.

\label{error}

\section{Distance of the LMC}

The Hipparcos cepheid distance scale can be
compared to that obtained from other methods, taking the
distance of the LMC as a point of comparison. Hipparcos cepheids give :
\begin{equation}
\mu_{LMC}=18.70 \pm 0.16\ [-0.03]\ \  {\rm \small(Hiparcos\ parallaxes,\ modified\ from\
FC)}
\label{lmc}
\end{equation}

The other two main calibrations of the cepheid PL zero-point are the surface
brightness technique and cepheids in clusters and associations. From the surface brightness
method, Gieren, Fouqué \& Gomez (1998) obtain:
\[
\mu_{LMC}= 18.46\, \pm 0.02\,\, [+ 0.06]
\]
where the 0.02 term is the internal statistical uncertainty, and the [+0.06] term is a
metallicity correction that the authors chose not to implement. In order
to put this value on the same scale as~(6), we force the slope of the PL
relation to $\delta\equiv-$2.81 as in FC, which yields $\mu_{LMC}=18.52$ (P. Fouqué,
priv. comm.). In the absence of an external error budget we increase the
uncertainty to 0.1 mag :
\[
\mu_{LMC}= 18.52 \pm 0.1 ?\ [+0.06] \\  {\rm \small (surface\ brigthness,\ modified\ from\
Gieren\ et\ al.\ 1998)}
\]
A recent cluster cepheid calibration is Laney \& Stobie  (1994) who find :
\[
\mu_{LMC}= 18.49 \pm 0.04\,\, [+ 0.04]
\]
where the last term is a metallicity correction. We adjust this result to the
new Hipparcos Hyades parallax of $\mu$=3.33 :
\[
\mu_{LMC}= 18.55 \pm 0.04\ [+ 0.04]\\  {\rm \small (cluster\ cepheids,\ modified\ from\
Laney\ \&\ Stobie\ 1994)}
\]
It is noticeable that the final agreement between the three different
calibration is compatible with the statistical
uncertainties. A weighted mean, applying half the systematic corrections
given in brackets, yields for the distance modulus of the LMC :
\[
\mu_{LMC}= 18.58 \pm 0.05
\]

We conclude that while the Hipparcos parallax calibration does indicate that the PL zero-point may be at the bright end of previous uncertainty intervals, it is not incompatible with other determinations. Or, depending on our viewpoint, that while the Hipparcos PL zero-point is within one sigma of previous calibrations, it does indicate that the PL zero-point is near the bright end of previous uncertainties.

\section{A note on kinematical determinations}

\label{kin}

Feast et al. (1998) have calculated the cepheid PL zero-point $\rho$ from kinematical data, by comparing Hipparcos proper motions and radial velocities. They adjust $\rho$ in order to obtain the same value for the Oort constant A from proper motions and from radial velocities. They find that this approach favours a brighter zero-point similar to \FC. A brighter zero-point would also decrease the mismatch between the rotation curve of cepheids and HII regions in the outer disc (see Pont et al. 1997). One could then conclude that the kinematics favours a longer distance scale.
It is striking however that the ``LM method'', also based on kinematics, finds such a faint zero-point for cepheids (see Par.~8). In addition, the RR Lyrae
statistical parallax analyses (such as Fernley et al. 1998), also kinematical,
 yield a much fainter distance modulus for the LMC
($\mu\sim18.3$). This could lead us to wonder whether all assumptions implicit in the kinematical methods about the cepheid or RR Lyrae velocity field  are really fulfilled. If not, it could turn out to be the key to understanding the puzzling disagreement
between the Hipparcos  cepheid zero-point on one hand, the LM
method for cepheids and RR Lyrae statistical parallaxes on the other.

\acknowledgments

This study has benefited a lot from enlightening discussions with Laurent Eyer, Michael Feast, Pascal Fouqué and Xavier Luri. I also thank Martin Groenewegen and Barry Madore for useful exchanges.

\pagebreak

\end{document}